\documentclass[12pt]{iopart}

\usepackage{epsfig}

\begin{document}

\title{Andreev bound states for cake shape superconducting-normal systems}
\author{ J.~Cserti\dag, B.~B{\'e}ri\dag, P.~Pollner\dag,
and Z. Kaufmann\dag}
\address{\dag\ Department of Physics of Complex Systems, E{\"o}tv{\"o}s
University, H--1117 Budapest, P\'azm\'any P{\'e}ter s{\'e}t\'any 1/A,
Hungary}

\ead{cserti@complex.elte.hu}

%\date{\today}

\begin{abstract}

The energy spectrum of cake shape normal - superconducting systems is
calculated by solving the Bogoliubov-de Gennes equation. 
We take into account the mismatch in the effective masses and Fermi
energies of the normal and superconducting regions as well as the
potential barrier at the interface. In the case of a perfect interface and
without mismatch, the energy levels are treated by semi-classics. 
Analytical expressions for the density of states and its integral, the step
function,  are derived and compared with that obtained from exact
numerics. 
We find a very good agreement between the two calculations.
It is shown that the spectrum possesses an energy gap and 
the density of states is singular at the edge of the gap. 
The effect of the mismatch and the potential barrier on the gap is also
investigated.  
\end{abstract}


\submitto{\JPC}

\maketitle

\section{Introduction}

At the interface of the mesoscopic hybrid normal -- superconducting 
(NS) junctions the underlying physics is controlled by 
the coherent evolution  of
electrons into holes due to the mechanism called 
Andreev reflection~\cite{Andreev}.  The growing interest in studying the 
role of the phase coherent Andreev reflection is motivated by the recent
technological progress in manufacturing almost ballistic semiconductors of
mesoscopic size coupled to a superconductor 
(for  overview of the recent progress in this field see e.g.\  
Ref.~\cite{Carlo-konyv,Schon1,Carlo1,Colin1}).
A ballistic normal dot weakly coupled to a superconductor is commonly
called an Andreev billiard. The excitation spectrum (Andreev states) of 
such NS systems can be calculated from 
the Bogoliubov-de Gennes equation~\cite{BdG-eq} (BdG).  
A film of normal metal in contact with semi-infinite superconductor was 
considered by P.\ G.\ de~Gennes and 
D.\ Saint-James~\cite{deGennes-Saint-James}, can be regarded as the first 
such Andreev billiard.  
The bound states of Andreev billiards have been extensively studied
over the past ten 
years~\cite{Kosztin,Melsen,Melsen-2,Altland,Lesovik,Nazarov,Heny,Richter1,Goldbart:cikk,box_disk:cikk,Gap_cikk,LogSq_cikk}.
The excitation spectrum were studied for SNS 
junctions~\cite{Kulik_Bardeen,Carlo_2_3,Bagwell,Gyorffy,Zagoskin-1:cikk},
too.
The energy levels of the strong magnetic field were investigated   
for a semi-infinite N region in contact with a semi-infinite 
S region~\cite{Ulrich} and  ring--shaped Andreev billiards~\cite{Bdisk:cikk}. 
The Andreev bound states for superconducting-ferromagnetic systems  
have also been studied 
(see e.g., Ref.~\cite{Ferro:cikk} and refs.~therein).

An important physical quantity for Andreev billiards is the density of
states (DOS). For perfect NS interface (in the absence of tunnel barrier
at the NS interface, and the mismatch in the Fermi energies 
and effective masses between the N and S regions)
a semi-classical Bohr-Sommerfeld approximation of the DOS was derived for 
excitation energies close to the 
Fermi energy~\cite{Melsen,Melsen-2,Nazarov,Heny,Richter1}.   
It was shown that the DOS can be related to the purely 
geometry-dependent path length distribution $P(s)$ which is 
the classical probability that an electron entering the N region at 
the NS interface returns to the interface after a path length $s$. 
In recent works~\cite{Gap_cikk,LogSq_cikk,Ferro:cikk}
an improved semi-classical Bohr-Sommerfeld approximation of the DOS has been
presented, in which the energy dependent phase shift for Andreev reflections 
is properly taken into account.
It has been shown that the exact quantum mechanical calculations 
for different Andreev billiards with perfect NS interface gives good agreement
with the semi-classical predictions for all energies below the bulk 
superconducting gap (not only at energies close to the Fermi energy). 
For a NS system possessing a finite cutoff in $P(s)$, a gap appears in
the energy spectrum which can be as large as half of 
the bulk superconducting gap~\cite{Gap_cikk}.   
Based on the semi-classical expression of the DOS, a simple formula 
for this gap was also derived for such Andreev billiards.

\begin{figure}[hbt]
\centerline{\epsfig{figure=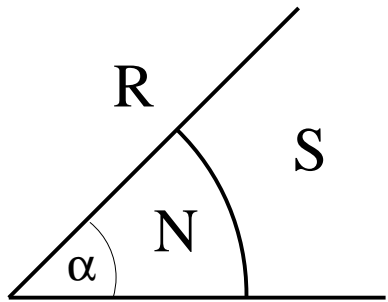,height=4cm}}
\caption{A cake shape Andreev billiard formed from a cake shape of normal
dot (N) in contact with a superconductor (S).
\label{geometria:fig}}
\end{figure}

In this paper our aim is twofold. First, we propose a new class of Andreev
billiards which can also exhibit a gap owing to the finite cutoff in
$P(s)$. Second, we study how the gap is suppressed for realistic NS
systems in which the NS interface is non-perfect. 
To this end we consider a `cake shape Andreev billiard' shown in
Fig.~\ref{geometria:fig} in which the boundaries of the N and S regions
are formed by two radii at an angle $\alpha$ 
and the NS interface is an arc between them.  

Recently Adagideli et al.~\cite{Goldbart:cikk} investigated 
the Andreev billiards formed from an arbitrary shape of normal region 
surrounded by superconductors. Their theoretical predictions have been
tested numerically in the case of circular shape of normal region.  
Similarly, the circular case has been studied by Stone~\cite{Stone:cikk}.  
This NS system is very similar to our cake shape Andreev billiards 
for $\alpha = 2\pi$. The only difference is that in our case there is
an infinite potential barrier along a radius.

The text is organized as follows. In Sec.~\ref{intro:sec} a quantization
condition (secular equation) is derived from the BdG equation for 
the general case of non-perfect NS interface. 
In Sec.~\ref{perfect_semi:sec} the density of states is derived from
semiclassics for the case of perfect NS interface. 
Our numerical results for the gaps in case of non-perfect NS interface 
are presented in Secs.~\ref{Z:sec} and~\ref{mismatch:sec}. 
Finally, the conclusions are given in Sec.~\ref{conclusion:sec}.

\section{Exact quantum calculation} \label{intro:sec}

In this section we derive a secular equation for 
the energy levels of cake shape Andreev billiards shown in
Fig.~\ref{geometria:fig}, in the general case of non-perfect NS interface.  
The BdG equation for the NS systems can be written as
\begin{equation}
\left(\begin{array}{cc}
H_0  & \Delta({\bf r}) \\ 
\Delta({\bf r})^* & - H_0 
\end{array}    
 \right)
\Psi
= E \,  \Psi,
\end{equation}
where $H_0 = {\bf p}^2/2m_{\rm eff} + V({\bf r})- \mu$ is 
the single-particle Hamiltonian,    
$\mu = E_{\rm F}^{(N)}, E_{\rm F}^{(S)}$ are the Fermi energies,  
$m_{\rm eff}= m_{\rm N}, m_{\rm S}$ are the effective masses 
in the N/S regions, $\Psi$ is a two-component wave function,  
$E$ is the quasi-particle energy measured from the Fermi energy 
$E_{\rm F}^{(N)}$.
The tunnel barrier $V({\bf r})$ at the NS interface in polar coordinates
$(r, \varphi)$ is modelled in a usual way 
by $V(r,\varphi)=U_0 \, \delta(r-R)$.  
We also adopt the usual step-function model~\cite{BTK} 
for the pair potential and take $\Delta({\bf r})= \Delta \Theta (r-R)$.
An infinite potential is assumed at the straight segments of the boundary
of the NS system, i.e., the Dirichlet boundary conditions are  
applied for the wave function $\Psi$. 
In this work we are interested in the discrete energy levels, i.e., it is
assumed that $0 < E < \Delta$.  

It is easy to see that the Hamiltonian is separable in polar coordinates.  
The $\varphi$ dependence of wave function $\Psi$ is identical 
in the N and S regions.  
After separating the $\varphi$ dependence of $\Psi$ 
the BdG equation reduces to two Bessel equations~\cite{Abramowitz} 
for the radial dependence of the two components of $\Psi$.
Thus, one can show that the ansatz for the wave functions 
in the N region can be written as 
\label{hullfv:e}
\begin{equation}
\Psi^{\rm{(N)}}_{m} (r,\varphi) =  
\left( \begin{array}{l} 
a_+  J_{\nu_m}(k_+ r)  \\[1ex]
a_-  J_{\nu_m}(k_- r) 
\end{array} \right) \, \sin \left(\nu_m \varphi \right), 
\label{hullfvN}
\end{equation}
while in the S region the wave function has the form 
\begin{equation}
\Psi^{\rm{(S)}}_{m} (r,\varphi) = \left[
c_+ \left( \begin{array}{l} 
\gamma_+  \\[1ex] 1  
\end{array} \right) 
H_{\nu_m}^{\left(1 \right)}(q_+ r) +  
c_- \left( \begin{array}{l} 
\gamma_-  \\[1ex] 1  
\end{array} \right)
H_{\nu_m}^{\left(2 \right)}(q_- r) 
\right]
\, \sin \left( \nu_m\varphi \right), 
\label{hullfvS}
\end{equation}
where $\nu_m = \frac{m\pi}{\alpha}$ ($m=1,2,\cdots$),  
$J_{\nu_m}(x) $ and $H_{\nu_m}^{\left(1,2 \right)}(x)$ are the Bessel and
the Hankel functions~\cite{Abramowitz}, and 
\begin{eqnarray}
k_{\pm}  &=& k_{\rm{F}}^{\rm{(N)}}
\sqrt{1 \pm \frac{E}{E_{\rm{F}}^{\rm{(N)}}}},   \\
q_{\pm}  &=& k_{\rm{F}}^{\rm{(S)}}
\sqrt{1 \pm i\,\frac{\sqrt{\Delta^2 - E^2 }}{E_{\rm{F}}^{\rm{(S)}}}}, \\
\gamma_{\pm} &=& e^{\pm i \arccos \left(E/\Delta \right)}. 
\end{eqnarray}
Here $+/-$ refer to the electron/hole like quasi-particle 
excitation and the Fermi wave numbers in the N/S regions are 
given by $k_{\rm{F}}^{\rm{(N)}} = 
\sqrt{2 m_{\rm N}  E_{\rm{F}}^{\rm{(N)}}/ \hbar^2}$ and 
$k_{\rm{F}}^{\rm{(S)}} = 
\sqrt{2 m_{\rm S}  E_{\rm{F}}^{\rm{(S)}}/ \hbar^2}$.  
The eigenfunctions of the energy levels are labeled by a fixed integer
number $m$.  
In the N region only the Bessel function can be used for the radial 
wave function since the Neumann function is singular at the origin.   
The wave function $\Psi$ in the S region must tend to zero as $r \to \infty$. 
This condition can be satisfied by choosing the appropriate Hankel function 
for the electronic and hole like component of $\Psi$ in the following way.
It is known~\cite{Abramowitz} that for large $r$ the Hankel functions  
$H_{\nu_m}^{\left(1\right)}(q_+ r) \sim \sqrt{2/(q_+r)}\, e^{iq_+ r}$. 
Since $q_+$ has a positive imaginary part one finds that 
$H_{\nu_m}^{\left(1\right)}(q_+ r) \to 0$ for $r \to \infty$. 
Similarly,  $H_{\nu_m}^{\left(2\right)}(q_- r) \to 0$ for $r \to \infty$.  

The $\varphi$ dependent part of the wave functions in the N and S regions 
ensures that they satisfy the Dirichlet boundary conditions at the
straight segments of the Andreev billiard.   
At the NS interface the matching conditions~\cite{Kato1,box_disk:cikk} 
should be applied.
The four coefficients $a_{\pm},  c_{\pm}$ in Eqs.~(\ref{hullfvN}) and
(\ref{hullfvS}) are determined from the matching conditions:
\begin{eqnarray}
\Psi^{\rm{(N)}}_m 
\rule[-1.6ex]{.2pt}{4ex}\;\raisebox{-1.5ex}{$\scriptstyle r=R$}
& = &  \Psi^{\rm{(S)}}_m 
\rule[-1.6ex]{.2pt}{4ex}\;\raisebox{-1.5ex}{$\scriptstyle r=R$} \,\, , \\[2ex]
\frac{d}{d r} \left[\Psi^{\rm{(N)}}_m 
- \frac{m_{\rm{N}}}{m_{\rm{S}}}\Psi^{\rm{(S)}}_m \right]
\rule[-1.6ex]{.2pt}{4ex}\;\raisebox{-1.5ex}{$\scriptstyle r=R$}  
& = & - \frac{2m_{\rm{N}}}{\hbar^2} U_0 \Psi^{\rm{(S)}}_m 
\rule[-1.6ex]{.2pt}{4ex}\;\raisebox{-1.5ex}{$\scriptstyle
r=R$}.
\end{eqnarray}
\label{matching:eq}
Substituting the ansatz given by Eqs.~(\ref{hullfvN}) and (\ref{hullfvS})
into Eq.~(\ref{matching:eq})  
one finds the following secular equation for the eigenvalues 
$E$  of the NS system for fixed mode index $m$: 
\begin{equation} 
\rm{Im} \left \{\gamma_+ D^{\rm{(+)}}_{m}(E) \, 
D^{\rm{(-)}}_{m}(E) \right \}=0,
\label{DNS}
\end{equation}
where ${\rm Im}\{. \}$ stands for the imaginary part and 
the two by two determinants $D^{\rm{(\pm)}}_m(E)$ are given by  
\begin{eqnarray}
D^{\rm{(+)}}_m(E) &=&  
\left| \begin{array}{cc}
D^{\rm{(+)}}_{11}  & D^{\rm{(+)}}_{12} \\
D^{\rm{(+)}}_{21}  & D^{\rm{(+)}}_{22} 
\end{array} \right|, 
\label{De-disk} 
\\[1ex]  
D^{\rm{(-)}}_{m}(E) &=& 
{\left[D^{\rm{(+)}}_m(-E)\right]}^* , 
\end{eqnarray}
with matrix elements 
\begin{eqnarray}
D^{\rm{(+)}}_{11}   &=&  J_{\nu_m}(k_+ R),  \\
D^{\rm{(+)}}_{12}   &=&  H_{\nu_m}^{\left(1 \right)}( q_+R), \\
D^{\rm{(+)}}_{21}   &=&  k_+ J_{\nu_m}^\prime(k_+ R),  \\ 
D^{\rm{(+)}}_{22}   &=& \!\!\!
- 2 k_{\rm{F}}^{\rm{(N)}} Z H_{\nu_m}^{\left(1 \right)}(q_+ R)  
\! + \! \frac{m_{\rm{N}}}{m_{\rm{S}}} q_+ 
H_{\nu_m}^{\left(1 \right)\prime}(q_+ R), 
\end{eqnarray}
\label{De}
and $Z=U_0 k_{\rm{F}}^{\rm{(N)}}/(2 E_{\rm{F}}^{\rm{(N)}}) $ 
is the normalized barrier strength used 
by Blonder et al.~\cite{BTK}. 
The primes denote the derivatives of the Bessel and Hankel functions
with respect to their arguments.  
For a given quantum number $m$ the energy levels of  cake shape Andreev
billiards for non-perfect interface are the solutions of the
secular equation given by Eq.~(\ref{DNS}). 
This equation is exact in the sense that the usual Andreev approximation
is not used. In the Andreev approximation it is assumed that 
$\Delta/E_{\rm{F}}^{\rm{(S)}} \ll 1$ and quasiparticles whose
incident/reflected directions are approximately perpendicular to the NS
interface~\cite{Colin1}.    

\section{Perfect NS interface} \label{perfect_semi:sec}

In the case of a perfect NS interface, the semi-classical expression 
for the DOS can be derived from the secular equation (\ref{DNS}).  
We assume that there is no mismatch and tunnel barrier at the NS
interface ($m_{\rm N} =m_{\rm S}$, $E_{\rm{F}}^{\rm{(N)}}
=E_{\rm{F}}^{\rm{(S)}}$ and $Z=0$).
For simplicity, we shall omit the superscript N and S in the wave
numbers and the Fermi energies.

In the Andreev approximation 
($\Delta/E_{\rm{F}} \ll 1$) we take 
$k_+ \approx q_+$ in places where they are multiplied 
by the Bessel or Hankel functions in $D^{(+)}_m(E)$ 
given by Eq.~(\ref{De-disk}) but 
in the arguments of the Bessel and Hankel functions 
we keep them to be different.
To approximate $D^{(+)}_m(E)$ the Debye asymptotic 
expansion~\cite{Abramowitz} of the Bessel functions 
will be used  for $|\nu_m| < k_+ R-\sqrt[3]{k_+ R}$.
For Hankel functions  of a complex argument $z$ the principal asymptotic
forms~\cite{Abramowitz}  
$H_{\nu}^{\left(1 \right)}(z) 
\approx \sqrt{2/(\pi z)}\, e^{i\left( z-\frac{1}{2}\nu \pi 
-\frac{\pi}{4} \right)}$ for $|z| \to \infty$ will be applied. 
After some algebra, the quantization condition (\ref{DNS}) 
can be simplified as 
\begin{equation}
\frac{\vartheta(k_+ R, \nu_m) - \vartheta(k_- R, \nu_m) 
- \arccos \frac{E}{\Delta}}{\pi} = n,
\label{BS_E_m:eq} 
\end{equation} 
where  
\begin{equation}
\vartheta(x,\nu)  =  \sqrt{x^2-\nu^2} - 
\left|\nu \right| \arccos \frac{\left| \nu \right|}{x}, 
\end{equation}
and $n$ is an non-negative integer. 
This is the Bohr-Sommerfeld quantization condition for cake shape
Andreev billiards. The function $\vartheta(k_\pm R,\nu_m)$ is 
proportional to the radial action of the electron/hole like
quasiparticles~\cite{Brack}. For a given $m$ and $n$ the solution of
(\ref{BS_E_m:eq}) gives the $E_{mn}$ energy level in Bohr-Sommerfeld
approximation. 

We now introduce the so-called step function $N(E)$ 
which is the number of energy levels below the energy $E$. 
This is indeed the integrated DOS $\varrho(E)$ of the Andreev billiards: 
\begin{equation}
N(E) = \int_{-\infty}^E \, \varrho(E^\prime) \, dE^\prime = 
\sum_{mn} \, \Theta (E-E_{mn}), 
\label{N_E-def:eq}
\end{equation}
where $\Theta(x)$ is the Heaviside  function.   
To calculate the step function in Bohr-Sommerfeld approximation, 
 Eq.~(\ref{BS_E_m:eq}) is Taylor expanded in terms of the small quantity 
$E/E_{\rm F}$. Keeping the leading terms one finds
\begin{equation}
\frac{E}{E_{\rm F}}\, \sqrt{{\left(k_{\rm F}R\right)}^2 - \nu_m^2} = 
n\pi + \arccos\frac{E}{\Delta}.
\label{E_m-Taylor:eq}
\end{equation}   
Substituting the solutions $E_{mn}$ of this equation into 
(\ref{N_E-def:eq}) gives the step function $N_{\rm BS}(E)$ 
in Bohr-Sommerfeld approximation. 
Applying the Poisson summation formula~\cite{Berry-1,Brack} 
in the sum over $m$ of $N(E)$ and keeping only the non-oscillating
term we find 
\begin{equation}
N_{\rm BS}(E) = \sum_{n=0}^\infty \, 
\int_{-\frac{1}{2}}^{M+\frac{1}{2}} \, \Theta (E-E_{mn})\, dm 
= \sum_{n=0}^\infty \, m^* (E,n),
\end{equation}
where $m^*(E,n)$  is the solution of Eq.~(\ref{E_m-Taylor:eq}) for $m$  at
a given $E$ and $n$, 
and $M = \frac{\alpha}{\pi}\, k_{\rm F}R$ is 
the largest $m$ for which the argument of the square root 
in the left hand side of (\ref{E_m-Taylor:eq}) is still positive. 

\begin{figure}[hbt]
\centerline{\epsfig{figure=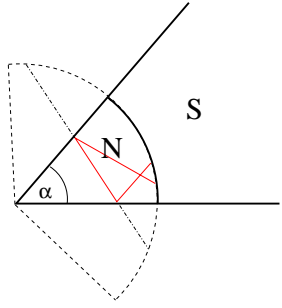,height=4cm}}
\caption{Unfolded trajectory in the cake shape Andreev billiard. 
\label{unfold:fig}}
\end{figure}

After a simple algebra we obtain the final form of the step function in
Bohr-Sommerfeld approximation:
\begin{equation}
N_{\rm BS} (E) = M \, \sum_{n=0}^\infty \, \left\{1- F[s_n(E)]\right\}, 
\label{BS_Stepfn:eq}
\end{equation}
where 
\begin{eqnarray}
F(s) &=& \int_0^s \, P(s^\prime) \, ds^\prime , 
\label{F_s:eq} \\
P(s) &=& \frac{1}{{\left( 2R \right)}^2} \, 
\frac{s}{\sqrt{1-{\left(s/2R\right)}^2}} \, \Theta(2R-s), 
\label{P_s:eq}   \\ 
s_n(E) &=& \frac{n\pi +\arccos \frac{E}{\Delta}}{E/\Delta} \, \pi \xi_c,
\label{s_n:eq}
\end{eqnarray}
$M= \frac{\alpha}{\pi}\, k_{\rm F}R$ and 
$\xi_c = \hbar v_{\rm F}/(\pi \Delta) = 2E_{\rm F}/(\pi k_{\rm F}\Delta)$ 
is the coherence length.  
It can be shown that $P(s)$ is the classical probability
that an electron entering the billiard at the NS interface
returns to the interface after a path length $s$. 
One can see that $P(s)$ depends only on the radius $R$ of the N region but
it is independent of the angle $\alpha$. 
This feature can be understood from a classical point of view. 
The unfolded trajectory of the particle starting and ending at the NS 
interface shown in Fig.~\ref{unfold:fig} is
always a chord of a circle of radius $R$, thus its length is independent of
the angle $\alpha$. 
The path length distribution $P(s)$ is normalized to
one, i.e., $\int_0^\infty\, P(s)\, ds =~1$. 
The integrated path length distribution is 
$F(s) = \left(1-\sqrt{1-{\left(s/2R\right)}^2}\, \right)\, \Theta(2R-s)$.

The DOS $\varrho_{\rm BS} (E) = dN_{\rm BS}(E)/dE$ in Bohr-Sommerfeld
approximation can easily obtained 
from (\ref{BS_Stepfn:eq}) and reads as  
\begin{eqnarray}
\varrho_{\rm BS} (E) &=& M\int_0^{\infty}\!\mbox{d}s\,P(s)
\left[\frac{s}{\hbar v_F} + \frac{1}{\sqrt{\Delta^2-E^2}}\right]
\nonumber \\
&\times& \sum_{n=0}^{\infty}\, \delta\left(\frac{s\,E}{\hbar v_F}-
\left(n\pi+\arccos\frac{E}{\Delta}\right) \right) \! ,
\label{DOS_BS:eq}
\end{eqnarray}
where $v_{\rm F}$ is the Fermi velocity.
Note that expressions (\ref{BS_Stepfn:eq}) and (\ref{DOS_BS:eq}) 
for the step function and the DOS in the semi-classical approximation 
are the same as that presented in Ref.~\cite{Gap_cikk,LogSq_cikk}
for general shapes of Andreev billiards. 
Our formula for the DOS, valid for all energy $E<\Delta$,   
can be regarded as an extension of that derived 
by Melsen et al.~\cite{Melsen,Melsen-2}, 
Lodder and Nazarov~\cite{Nazarov}, 
Schomerus and Beenakker~\cite{Heny}, and Ihra et al.~\cite{Richter1} 
in the limit $E \ll \Delta$. 
In these works, the  energy dependent phase shift 
$-\arccos(E/\Delta)$, due to Andreev reflections, was approximated 
by $\pi/2$, while in our result this phase shift is fully incorporated.  

\begin{figure}[hbt]
\centerline{\epsfig{figure=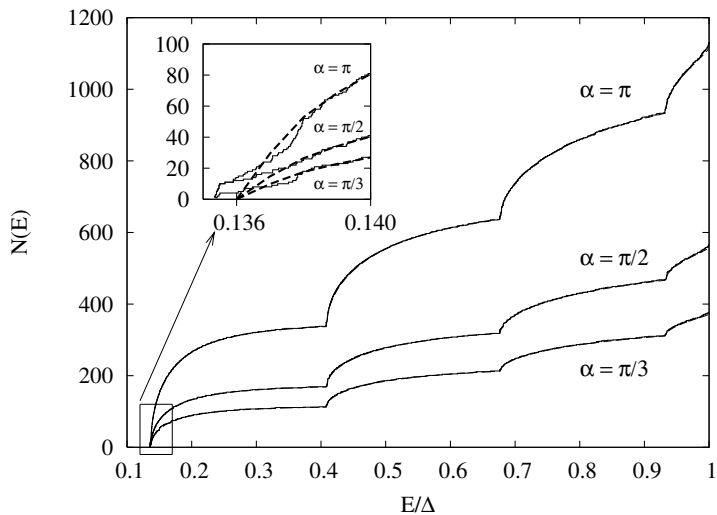,height=7cm}}
\caption{The exact step function $N(E)$ (solid lines) from
Eq.~(\ref{DNS}) and its semi-classical approximation 
$N_{\rm BS} (E)$ (dashed lines) given by Eq.~(\ref{BS_Stepfn:eq}) 
for the cake shape Andreev billiards of 
angles $\alpha = \pi, \pi/2, \pi/3$ as functions of $E/\Delta$. 
The inset shows an enlarged part of the main frame. 
The parameters are $k_{\rm F}R = 350$ and $\Delta/E_{\rm F} = 0.03$. 
With these data $\xi_c /R = 0.06 $. 
The numerically obtained exact gap from Eq.~(\ref{DNS}) is 
$E_{\rm gap}/\Delta = 0.1355$,  
while its semi-classical value from Eq.~(\ref{E_gap:eq}) is 
$E_{\rm gap}/\Delta = 0.1366$.
\label{Stepfn_perfect:fig}}
\end{figure}

In Fig.~\ref{Stepfn_perfect:fig} the numerically calculated exact step 
function $N(E)$ obtained by solving Eq.~(\ref{DNS}) 
together with its semi-classical approximation $N_{\rm BS} (E)$ 
from Eq.~(\ref{BS_Stepfn:eq}) as functions of the energy $E$ below the bulk
gap $\Delta$ are shown for angles $\alpha = \pi, \pi/2, \pi/3$. 
One can see that the agreement between the two results is excellent for
all energy $E<\Delta$. 
To see the difference between the two calculations, the inset shows 
an enlarged part of the main frame. 

The step functions shown in Fig.~\ref{Stepfn_perfect:fig} have cusps at some
energies. This implies that the DOS has peaks at these energy values. 
Since the semi-classical approximation of the exact step function is 
very good one can derive an expression for the positions of 
these peaks starting from (\ref{BS_Stepfn:eq}).   
It is clear from (\ref{P_s:eq}) that $P(s)$ is singular at $s = 2R$. 
This implies that the DOS will be singular at those energies 
when $s_n (E) =2R$. 
Using (\ref{s_n:eq}) and the approximation 
$\arccos \left(E/\Delta\right) \approx \pi/2 - E/\Delta$ 
the positions of these singularites in the DOS is given by
\begin{equation}
\frac{E_n^{\left(\rm sing \right)}}{\Delta} = 
\frac{\left(n+1/2 \right)\pi}{1+ 2R/(\pi \xi_c)}
\label{E_sing:eq}
\end{equation}
valid for all integral $n$ for which $E_n^{\left(\rm sing \right)} < \Delta$. 
It is worth mentioning that the positions of the singularities are
{\em independent} of the angle $\alpha$ of the cake shape Andreev
billiards. 

Figure~\ref{Stepfn_perfect:fig} also shows that in the energy spectrum of 
the system there is a quite large gap. 
This gap always exists for cake shape Andreev billiards. 
To derive a formula for the value of the gap one can also start from
the semi-classical approximation.    
It is obvious that $P(s)$ possesses an upper cut-off, i.e., $P(s) =0$
if $s>s_{\rm max} =2R$. 
Thus, from Eq.~(\ref{s_n:eq}) it follows that the energy spectrum has a
lower bound. An approximate value of this energy 
can be obtained by setting $n=0$ in (\ref{E_sing:eq}) since the position
of the singularity and the upper cut-off of $P(s)$ is identical 
for cake shape Andreev billiards. 
The lowest energy level (the gap of the spectrum) in semi-classical
approximation becomes 
\begin{equation}
\frac{E_{\rm gap}}{\Delta} = 
\frac{\pi/2}{1+ 2R/(\pi \xi_c)}. 
\label{E_gap:eq}
\end{equation}
For $\xi_c \ll R $ the gap is 
$E_{\rm gap}/\Delta \approx \pi^2 \xi_c/(4R)$.
It is clear that the value of the gap is also 
{\em independent} of the angle $\alpha$ of the cake shape Andreev
billiards. 
From Fig.~\ref{unfold:fig} one can see that the length of the longest 
possible chord is $2R$. 
Therefore, the cut-off of the path length distribution $P(s)$ is always 
$2R$ and obviously independent of $\alpha$.  

It is easy to express the gap $E_{\rm gap}$ in units of the 
mean level spacing  $\delta_N$ of the isolated normal region. 
It is well-known that $\delta_N = 2 \pi \hbar^2/(m_{\rm N} A)$, where 
$A= \frac{1}{2}\, R^2 \alpha$ is the area of the normal region~\cite{Brack}. 
Then, from (\ref{E_gap:eq}) one finds   
\begin{equation}
\frac{E_{\rm gap}}{\delta_N} = \frac{\alpha k_{\rm F} R}{16} 
\end{equation}
for  $\xi_c \ll R $. 
In a macroscopic sample $ k_{\rm F} R \sim R/\lambda_{\rm F} \gg 1 $,
where $\lambda_{\rm F}$ is the de Broglie wavelength. 
Thus, the gap in the cake shape Andreev billiards 
can be a large value on the energy scale of the mean level spacing.

\section{NS interface with potential barrier} \label{Z:sec}

\begin{figure}[hbt]
\centerline{\epsfig{figure=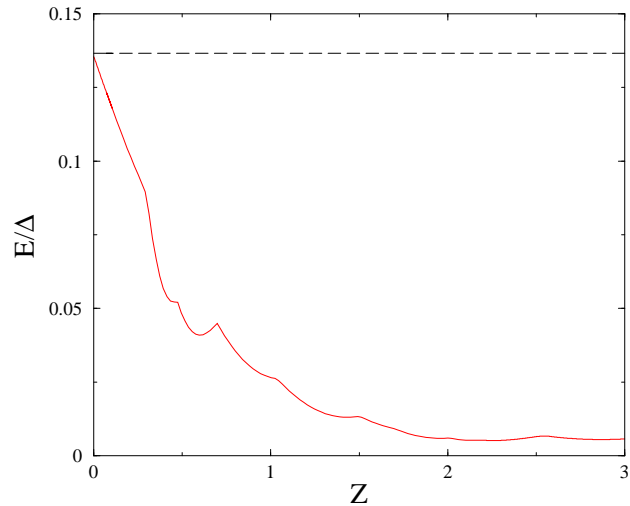,height=7cm}}
\caption{Numerically calculated energy gap  (in units of $\Delta$)  
as a function of $Z$. For $Z=0$ the semi-classical result 
from (\ref{E_gap:eq}) is indicated by the dashed curve. 
The parameters are the same as in Fig.~\ref{Stepfn_perfect:fig}, 
and $r_k=1$ and $r_v=1$.  
\label{Gap_z:fig}}
\end{figure}

In this section we investigate the dependence of the gap on the potential
barrier at the NS interface. 
Here we assume that the effective masses and the Fermi energies between 
the normal and the superconducting regions are the same. 
Following Mortensen et al.~\cite{Mortensen:cikk} the mismatch of the N and
S regions is characterized by the ratio of the Fermi wavenumbers and the
Fermi velocities of the two regions
$r_k = k^{(\rm N)}_{\rm F}/k^{(\rm S)}_{\rm F}$ 
and $r_v = v^{(\rm N)}_{\rm F}/v^{(\rm S)}_{\rm F}$. 
It is easy to show that the ratio of the effective masses 
and the Fermi energies appearing in the secular equation (\ref{DNS}) 
can be expressed by the parameters 
$r_k$ and $r_v$ in the following way 
$m_S / m_N  =  r_v / r_k$ and 
$E^{(\rm N)}_{\rm F}/E^{(\rm S)}_{\rm F}  =  r_k r_v $ .
The energy levels of the systems is calculated by solving the secular
equation (\ref{DNS}) for different $Z$, and for $r_k=1$ and $r_v=1$.
The result is shown in Fig.~\ref{Gap_z:fig}. 
It is seen from the figure that the gap is decreased with increasing
potential barrier. This is owing to the fact that the probability of the 
normal reflection is increased and the condition for perfect Andreev
reflection is supressed.

\section{NS interface in case of mismatch} \label{mismatch:sec}
\begin{figure}[hbt]
\centerline{\epsfig{figure=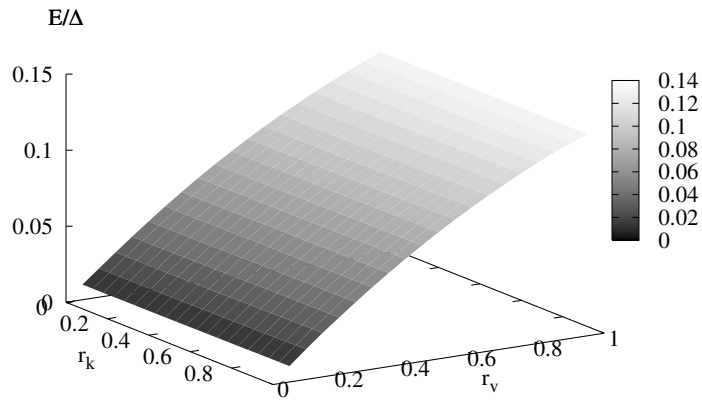,height=7cm}}
\caption{Numerically calculated energy gap (in units of $\Delta$) 
as a function of $r_k$ and $r_v$. 
The parameters are $\alpha = \pi $,  $k^{(\rm N)}_{\rm F} R = 350$,  
$\Delta / E^{(\rm N)}_{\rm F} = 0.03$ and $Z=0$.  
\label{mismatch:fig}}
\end{figure}

We now present our numerical results for the gap when there is a 
mismatch in the effective masses and the Fermi energies of 
the N and S regions. The numerical solution of the secular equation 
(\ref{DNS}) gives the energy levels of the system. The lowest level can be
associated to the gap. 
In Fig.~\ref{mismatch:fig} the gap is plotted as a function of 
$r_k$ and $r_v$. The surface plot shows that the suppression of the 
gap is very sensitive to the parameter $r_v$ 
but practically independent of $r_k$.

\section{Conclusions} \label{conclusion:sec}

We calculated the Andreev bound states of cake shape Andreev
billiards by solving the Bogoliubov-de Gennes equation. Matching the
wave functions in the normal and superconducting regions we 
obtained a secular equation for the energy levels
of the system. In the exact
calculation the different effective masses and Fermi energies of the two
regions, and the potential barrier at the NS interface are taken into
account. For a perfect interface we derived a semi-classical expression for
the step function and the DOS of the energy spectrum. These quantities are
expressed by the classical return probability $P(s)$ of the particles.  
It was found that the cake shape Andreev billiards always possesses a gap. 
An analytical expression was derived for the value of the gap 
for perfect interface.   
We also showed that the DOS has singularities at some energies given by a
simple formula. Since the cut-off of $P(s)$ is the same at which 
$P(s)$ is singular, we found that at the edge of the gap 
the DOS is singular. 
Similarly, the gap is independent of the angle $\alpha$ 
of the cake shape Andreev billiard. 
We think that these facts can be exploited experimentally to measure 
the value of the gap. 
The Andreev bound states determine the tunneling conductance of the NS 
systems. 
Thus, the measured conductance (proportional to the DOS)  
should be changed abruptly at the edge of the gap. 
A cake shape Andreev billiard with $\alpha = \pi $ can be made 
by cutting a semi-circle region from a semi-infinite bulk 
superconductor and replacing that by a normal region.       
    
We also investigated that how the non-perfect NS interface and mismatch in
the material parameters of the normal and superconducting regions can
affect the value of the gap. 
We found that the gap decreases with increasing potential barrier
at the NS interface. 
We also calculated the value of the gap as a function of 
the ratio $r_k$ of the Fermi wave numbers 
and the ratio $r_v$ of the Fermi velocities of the normal
and superconducting regions.  
In Ref.~\cite{Mortensen:cikk} it has been shown that 
$r_k$ determines the ratio of the sine of the incident and
outgoing angle of a particle at the NS interface. 
This implies, and it is indeed supported by our numerical results, 
that the value of the gap is very weakly dependent on the parameter
$r_k$. However, our numerics show a strong dependence of the gap 
on the parameter $r_v$: increasing the mismatch in the Fermi velocities 
decreases the value of the gap.

\ack 

This work is supported in part by the 
EU FP6 programme under Contract No. MRTN-CT-2003-504574,
the Hungarian-British TeT Grant No. GB-29/01, EPSRC, and 
the Hungarian Science Foundation OTKA Grant Nos. TO34832, D37788 
and T42981.
One of us (Z. K.)  thanks the Hungarian Academy of Sciences
for its financial support as a J\'anos Bolyai Scholarship.

%\end{references}

\end{document}